# Heisenberg exchange in the magnetic monoxides

Walter A. Harrison
Applied Physics Department
Stanford University
Stanford, CA 94305

Abstract

The superexchange intertacion in transition-metal oxides, proposed initially by Anderson in 1950, is treated using contemporary tight-binding theory and existing parameters. We find also a direct exchange for nearest-neighbor metal ions, larger by a factor of order five than the superexchange. This direct exchange arises from $V_{ddm}$ coupling, rather than overlap of atomic charge densities, a small overlap exchange contribution which we also estimate. For FeO and CoO there is also an important negative contribution, related to Stoner ferromagnetism, from the partially filled minority-spin band which broadens when ionic spins are aligned. The corresponding $J_1$ and $J_2$ parameters are calculated for MnO, FeO, CoO, and NiO. They give good accounts of the Néel and the Curie-Weiss temperatures, show appropriate trends, and give a reasonable account of their volume dependences. For MnO the predicted value for the magnetic susceptibility at the Néel temperature and the crystal distortion arising from the antiferromagnetic transition were reasonably well given. Application to $CuO_2$ planes in the cuprates gives $J=1220°K$, compared to an experimental 1500°K, and for $LiCrO_2$ gives $J_1=4\,50°K$ compared to an experimental 230°K.

## 1. Introduction

The interaction between electron magnetic moments on neighboring ions in solids is usually written in terms of Heisenberg exchange, $2J\mathbf{S}_i\cdot\mathbf{S}_j$ for the coupling between two spins, with angular momentum $\mathbf{S}_i$ and $\mathbf{S}_j$ in units of $\hbar$, (e. g., Kittel[1], p. 462, but here with positive $J$ for antiferromagnetism. The factor two is because the total is written $J\Sigma_{I,i}\,\mathbf{S}_i\cdot\mathbf{S}_j$ and each pair enters as $ij$ and $ji$.). The parameter $J$ is generally taken from experiment but Anderson[2] long ago proposed a *superexchange* mechanism for such coupling in ionic solids containing transition metals, based upon a coupling between the $d$ states on a magnetic ions and orbitals on the neighboring anions, $p$ states on oxygen in particular. The successive coupling of these oxygen orbitals with $d$ states on a neighboring magnetic ion produced a coupling between magnetic moments of fourth order in this *pd* coupling. He sought to estimate the magnitudes of the coupling[3] using rather crude estimates for the quantities which enter. More recently we[4] have given values for all of the parameters which enter this superexchange mechanism, based upon elementary descriptions of the electronic structure and aimed at other bonding and dielectric properties of these solids. In the present analysis we use these newer parameters to estimate the magnitude, and find the detailed form, of Anderson's superexchange interaction.

These parameters also include a direct coupling between *d* states on neighboring magnetic atoms, which gives rise to an additional coupling between moments, and we include this coupling also, calling it *direct exchange*. It is of a different physical origin than the "true direct exchange" introduced by Anderson[3], which was based upon Hartree-Fock exchange in the overlapping electron densities from the two magnetic ions, which we prefer to call *overlap exchange*. We estimate this contribution also and find that it is indeed small, as Anderson indicated. We shall find however that our direct exchange is considerably larger than superexchange.

The terminology is unfortunate because *exchange* has a clear meaning in Hartree-Fock theory and band theory: it is the lowering $U_x$ of repulsive Coulomb interaction between two electrons of the same spin on an atom because of the antisymmetry of their spatial wavefunction with respect to interchange, keeping electrons of the same spin from being close to each other. It is the origin of Hund's rule that the energy for an atom is lower if the spins of its electrons are aligned. The relative alignment of these total spins on neighboring atoms is quite a separate question. The strength of the superexchange, and our direct exchange, for the coupling between moments on different atoms would not be greatly changed (only through the lowering of the minority-spin *d* states) if the Hartree-Fock $U_x$ were taken to zero. Of the four contributions to the Heisenberg *J* we shall study, only the overlap exchange arises from this Hund's rule exchange.

We shall proceed in Section 2 by giving the parameters for the MnO electronic structure, the simplest case. We then proceed, treating the coupling between neighboring orbitals in perturbation theory, to calculate the total energy difference $\Delta E$ depending upon whether an Mn neighbor to a particular Mn ion has parallel rather than antiparallel spin alignment. This includes nearest-neighbor, $J_1$, superexchange and second-neighbor, $J_2$, superexchange and direct and overlap exchange $J_1$ between nearest neighbors. In Section 3 we give also the parameters for FeO, CoO and NiO, and indicate the needed changes in the calculation, and in Section 4 predict various magnetic properties and compare with experimental values where available.

## 2. Coupling between Moments in MnO

MnO, with all majority-spin states occupied and all minority-spin states empty, has the simplest electronic structure of the magnetic monoxides. The important states on the anion, oxygen, are the valence $p$ states, having energies given by the Hartree-Fock term value[5] –16.77 eV, approximately the removal energy of the corresponding electron from that atom (Appendix B). In the crystal the two $s$ electrons from the Mn atom have been transferred to these states so that all six states are occupied, as indicated in Fig. 1 (a). The Coulomb repulsion of these added electrons would raise this level, but that rise is approximately cancelled by the Madelung potential (as in the alkali halides[4]) so we do not change it.

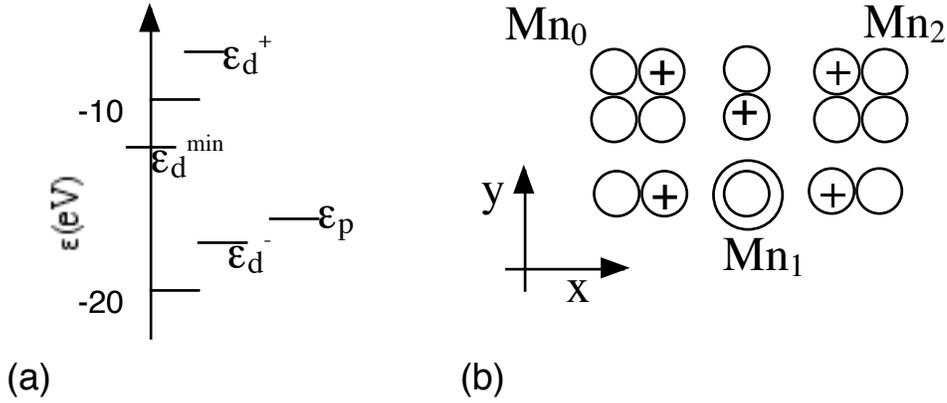

Fig. 1. In Part (a) are shown he essential energy levels in MnO, with five electrons per MnO in the Mn $\varepsilon_d^-$ state, six in the oxygen $\varepsilon_p$. The energy at which an electron could be added from another Mn is $\varepsilon_d^+$. The energy to flip the spin of a majority electron is $\varepsilon_d^{min}-\varepsilon_d^-$. In Part (b) is a segment of the lattice, showing a $d$ state $xy$ on an atom $Mn_0$, a $d$ state $3z^2-r^2$ on a nearest-neighbor $Mn_1$ and an $xy$ on a second-neighbor $Mn_2$. Also shown are oxygen $p$ states on two of the oxygen neighbors to $Mn_o$. The signs for the wavefunctions are for antibonding states.

The important states for the manganese ions are the $3d$ states, with Hartree-Fock term value –15.27 eV given in Ref. 4, but this value applies to a $d^5$ configuration with equal numbers of up and down-spin $d$ electrons. The exchange interaction $U_x = 0.78$ eV obtained from atomic spectra of manganese was listed in Ref. 4 (p. 589), so shifting 2.5 electrons to the majority spin leads to a term value for the majority spin of $\varepsilon_d^- = -17.22$ eV, approximately equal to the removal energy for these electrons, appropriate for spins fully aligned, and indicated in Fig. 1(a). Adding $2.5U_x$ to the Hartree-Fock term value gives $\varepsilon_d^{min}=-13.32$ eV which would be approximately the level to which a majority electron would need to be raised to flip its spin. That energy will not be of interest for MnO, where we will only partially populate these states from the neighboring Mn atoms. For that we must add to $\varepsilon_d^{min}$ the Coulomb repulsion of the majority electron which was *not* removed if we simply flipped the spin. We take that repulsion to be the $U_d$ = 5.6 eV (from Ref. 4, p. 645), leading to the $\varepsilon_d^+$ = −7.72 eV shown in Fig. 1 (a). The relevance of this choice of $U_d$ is discussed in Appendix B, and in Appendix A we look carefully at the correlation of electrons in multiple bonds to see that $\varepsilon_d^+$ obtained with this $U_d$ can be used directly in perturbation theory. These values will change when we go to FeO, CoO, and NiO, and at each step we add an additional minority electron at $\varepsilon_d^{min}$.

We next add the coupling between orbitals on neighboring ions, the largest being the coupling between the Mn $d$ states and the O $p$ states. For states with no angular momentum around the internuclear axis it is given by (Ref. 4, p. 643) $V_{pd\sigma} = -(3\sqrt{15}/2\pi)\hbar^2(r_d^3 r_p)^{1/2}/md^4$ and for one unit of angular momentum $V_{pd\pi} = - V_{pd\sigma}/\sqrt{3}$ . In Ref. 4, Table 15-1, two sets of values for the $r_d$ were listed, one obtained from Muffin-Tin Orbital (MTO) calculations by Andersen and Jepsen[6], and the other calculated directly and simply by Straub[7] from the atomic Hartree-Fock wavefunctions using the Atomic Surface Method (ASM). We indicated in Table 15-1 that the MTO values were to be preferred, because the calculations made fewer approximations, but listed the ASM values in the Solid State Table and have used them often since, perhaps because then all parameters were obtainable simply from free-atom states. The values generally differed by only a few percent, but for the elements of interest here (Mn, Fe, Co, and Ni), the MTO values are larger by about 16% and we discovered during the present study that they account much better for the magnetic properties (they enter to the sixth power, as we shall see). We therefore shall use the MTO values, and also check (at the end of the paper) the change which this would make in other recent calculations we have made for transition-metal systems. With $r_d$ = 0.925 Å for Mn (Ref. 4, p. 539) and $r_p$ = 4.41 Å for O (Ref. 4, p. 644; a more complete set for anions is given in Ref. 8), and a spacing of $d$ = 2.22 Å , this gives $V_{pd\sigma}$ = −1.084 eV and $V_{pd\pi}$ = 0.626 eV.

The couplings between Mn $d$ states, second neighbors at a distance $r=\sqrt{2}d=3.14$ Å, are much smaller. We use[4] $V_{dd\sigma} = -45\hbar^2 r_d^3/(\pi m d^5)$ for nearest-neighbor interactions, which would give −0.283 eV if we simply replaced $d$ by the larger $r$ . However, here nearest neighbors are oxygen and we certainly expect exponential variation at such large metal-metal spacings. It is preferable to fit this nearest-neighbor result to an exponential, giving a value smaller by a factor 0.71 ( equal to $\exp(-5(r-d)/d)/(d/r)^5)$ , $V_{dd\sigma}=-0.202$ eV. Then $V_{dd\pi}=-2V_{dd\sigma}/3$ and $V_{dd\delta}=V_{dd\sigma}/6$. These will enter our direct exchange interaction. These couplings broaden the levels into narrow bands, but with a wide gap between the occupied and empty states so that the system is insulating, and we will not need to be concerned with the structure of the bands. When we turn to FeO and CoO the minority-spin states will be partially occupied and we will need to consider them more carefully. Of more importance for MnO, the coupling will affect the relative energy of antiferromagnetic and ferromagnetic spin alignments, which is our concern here.

The total energy differences, in the context of self-consistent-field theory, can be taken as the difference in the sum of occupied-state eigenvalues, as long as the atomic charges are held constant. (e. g., Ref. 4, p. 15.) If we imagine a Hamiltonian matrix with the levels shown in Fig. 1 (a) on the diagonal, the addition of coupling in the off-diagonal elements does not change the sum of all eigenvalues and it will be simplest to calculate only the shift of the few empty levels, knowing that the total shift in occupied values will be equal and opposite.

We note also that the energy difference between these empty levels and the occupied ones is so large that perturbation theory should be adequate. That conjecture was confirmed, again in the context of self-consistent-field theory, by going beyond perturbation theory using the method applied earlier to dielectric properties of insulators[8]. In this approach, as would be appropriate here, three-by-three Hamiltonian matrices were obtained, based upon states at $\varepsilon_d^+$, $\varepsilon_d^-$, and $\varepsilon_p$, which if evaluated to fourth order in the couplings would be equivalent to perturbation theory. We also solved the three-by-three for MnO exactly and found that this reduced the shifts in the empty levels, and therefore the Heisenberg $J$'s, by only 10%. We further solved for the states themselves and calculated the self-consistent shifts in the diagonal elements, which reduced the estimated shifts and $J$'s by another 10%, principally from the upward shift of $\varepsilon_d^+$ due to charge transfer from the $p$ states to the empty $d$ states. We considered these shifts small enough that we did not complicate the analysis by including them. A more complete account of this approach, an Appendix C, is available from the author on request (*walt@stanford.edu*), but is not included here.

In the analysis here it will actually be important to go beyond self-consistent-field theory. In Appendix A we discuss a simplified problem with a single pair of identical states, coupled by a single *V* and containing two electrons, with an extra energy *U* if both electrons are on the same atom.

This problem is so limited, with only six two-electron states, that it can be solved exactly. It is seen that when $U$ is large, the effect on the total energy of $V$ is to lower the energy by $-4V^2/U$, while in self-consistent-field theory (Local-Density Approximation(LDA), or Unrestricted Hartree-Fock, for example) the lowering in energy is only $-2V^2/U$. One way of intuitively explaining the difference is to say that in LDA an electron localized on one atom can jump to the other, at higher energy by U and return, contributing $-V^2/U$ to the energy. In the full solution an additional, equal, contribution comes from one electron jumping to the other atom, but then the *other* electron jumps to the first atom. Adding the contributions from both electrons leads to the $-4V^2/U$.

Our use of perturbation theory will reduce the problem to the effects of coupling between specific pairs of electronic states, and for each such pair we shall use the $-4V^2/U$ contribution rather than the $-2V^2/U$ that would be obtained in self-consistent-field theory. Also by adding the effects of orbitals on specific pairs of neighbors, the resulting change in energy be represented by a Heisenberg parameters $J$. Thus in detail we shall calculate the shift in the empty (minority) $d$ states due to interactions involving a particular neighboring magnetic ion, obtaining $\Delta E$, the energy for the pair of ions if their spins are parallel minus the energy if they are antiparallel. Since our calculation of energies corresponds to taking these spins as aligned or antialigned along some axis ($\mathbf{S_1} \cdot \mathbf{S_2} = \pm(5/2)^2$ for Mn), the corresponding $J$ is given by

$$J = \Delta E/4S^2. \tag{1}$$

If $\Delta E$ is in eV, we may multiply by $1/k_B = 11604$ °K/eV to obtain traditional units of °K. [This energy difference $4JS^2 = 25J$ is not far from the $30J$ which would be obtained from full quantum treatment of the added angular momenta for an isolated pair of ions. Then one writes the total spin for the pair as $\mathbf{S}_p = \mathbf{S}_1 + \mathbf{S}_2$ so $\mathbf{S_1} \cdot \mathbf{S_2} = (S_p^2 - S_1^2 - S_2^2)/2$. With $S_p^2 = S_p(S_p+1)$, etc., the difference in energy between parallel spin, $S_p = 5$, and antiparallel spin, $S_p = 0$, is $5 \times 6J$. We shall use this full quantum form when calculating magnetic properties from our estimated $J$'s.]

We may see how this energy calculation proceeds for superexchange by considering a single empty $d$ state on a Mn ion such as the $xy$ orbital shown on $Mn_0$ in Fig. 1 (b), and considering the influence of the orbitals of the nearest-neighbor $Mn_1$. This $xy$ orbital is coupled to $p$ states on four of its neighboring oxygens by $V_{pd\pi}$ so that in perturbation theory this $xy$ orbital contains terms $V_{pd\pi}|p\rangle/(\varepsilon_d^+ - \varepsilon_p)$ from each of the two $p$ states shown. Of the orbitals $xy$, $yz$, $zx$, $x^2-y^2$, and $3z^2-r^2$ on the Mn$_1$ ion, only the $3z^2-r^2$ orbital shown is coupled to the combination of the two, to each by $-V_{pd\sigma}/2$. The terms from the two $p$ states add to give a second-order coupling $V=V_{pd\pi}V_{pd\sigma}/(\varepsilon_d^+ - \varepsilon_p)$ between the $d$ states on the two ions. The same coupling applies to the $3z^2-r^2$ orbital on Mn$_0$, coupled to an $xy$ orbital on Mn$_1$. Similarly, the $xz$ orbital on Mn$_0$ is coupled to a $yz$ orbital on Mn$_1$ by $V=V_{pd\pi}^2/(\varepsilon_d^+ - \varepsilon_p)$, as is the $yz$ orbital on Mn$_0$ to the $xz$ orbital on Mn$_1$. The $x^2-y^2$ orbital is coupled to $p$ states on both of these oxygens, but the combination is not coupled to any $d$ state on Mn$_1$.

Now if the spin on Mn$_1$ is opposite to that on Mn$_0$, the minority-spin states on Mn$_0$ are coupled to majority-spin states on Mn$_1$ and the sum of the energy shifts upward of the empty minority $d$ states added for the *two* ions is

$$\Delta E_1 = 4\sum \frac{V^2}{U} = 4\frac{2V_{pd\sigma}^2 V_{pd\pi}^2 + 2V_{pd\pi}^4}{(\varepsilon_d^+ - \varepsilon_p)^2(\varepsilon_d^+ - \varepsilon_d^-)}, \quad \text{(superexchange, MnO)} \quad (2)$$

where the sub-1 indicates nearest-neighbor Mn. On the other hand, if the spin on the Mn$_1$ states are parallel to those on Mn$_0$ it is only coupling between empty states and there is no net shift in energy of the empty (nor of the filled) states from these fourth-order terms, and the second-order terms are the same for both parallel and antiparallel cases. Thus this $\Delta E_1$ is exactly the $\Delta E$ determining the $J$ in Eq. (1).

The shift in the energy for superexchange coupling with second neighbors, between M$_0$ and Mn$_2$ in Fig. 2 (b), is still simpler. The $xy$ states on Mn$_0$ are coupled only to $xy$ states on Mn$_2$, and the $zx$ states on Mn$_0$ only to $zx$ states on Mn$_2$, and both matrix elements are $V_{pd\pi}$. The $3x^2-r^2$ states on the two ions are coupled with both matrix elements $V_{pd\sigma}$, and the $yz$ and $y^2-z^2$ states are *not* coupled, to give the energy for this coupled pair of

$$\Delta E_2 = 4\frac{V_{pd\sigma}^4 + 2V_{pd\pi}^4}{(\varepsilon_d^+ - \varepsilon_p)^2(\varepsilon_d^+ - \varepsilon_d^-)}. \quad \text{(superexchange, MnO)} \quad (3)$$

It may be interesting that we could similarly calculate the energy difference for an isolated set, Mn-O-Mn, as a function of the angle $\theta$ between the axes to obtain a $J(\theta)$. It would correspond to Eq. (2) for $\theta = \pi/2$, and we could double it to include the effect of both oxygens, but it would still differ in the term in $V_{pd\sigma}^2 V_{pd\pi}^2$ because of the interference between matrix elements for the two oxygens in the crystal (perhaps analogous to the adding which doubled the coupling in correlated pairs in Appendix A). It is better to define the $J$'s in terms of the full crystal structure, as in Eqs. (2) and (3).

Noting that $V_{pd\pi}^2 = V_{pd\sigma}^2/3$ we may obtain the ratio $J_2/J_1 = \Delta E_2 / \Delta E_1 = 11/8 = 1.38$ for superexchange. Substituting the values given above gives $\Delta E_1 = 6.32$ meV and $\Delta E_2 = 8.68$ meV, corresponding to (Eq. (1)) $J_1 = 2.94$ °K and $J_2 = 4.04$°K. These are much smaller than values derived by Lines and Jones[9] from experiment.

We noted above however that there is also a direct coupling between $d$ states on neighboring ions, corresponding to $V_{dd\sigma}$, $V_{dd\pi}$, and $V_{dd\delta}$. These apply to nearest neighbors Mn ions and give quite directly

$$\Delta E_1 = 4 \frac{V_{dd\sigma}^2 + 2V_{dd\pi}^2 + 2V_{dd\delta}^2}{\varepsilon_d^+ - \varepsilon_d^-} \quad , \text{ (direct exchange, MnO)} \quad (4)$$

corresponding to $\Delta E_1 = 33.4$ meV, or 15.6°K, raising $J_1$ to 18.4°K.

We may also estimate the overlap exchange, mentioned at the beginning. The normalized probability density for the atomic $d$ states is approximately $\rho(r) = (\mu^3/\pi)\exp(-2\mu r)$, with $\hbar^2\mu^2/2m = -\varepsilon_d$ (e. g., Ref. 4 p. 355. We use the Hartree-Fock term value $\varepsilon_d$. Note also that orthogonalizing a $d$ state on one atom to that on another, $|d_1\rangle \to |d_1\rangle - |d_2\rangle\langle d_2|d_1\rangle$ adds $-\langle d_2|d_1\rangle^2 U_x$ for the second atom, but subtracts it from the first, making no additional change in energy.). We let the nuclei be separated by a distance $\mathbf{s}$, equal to $d\sqrt{2}$ here. We may calculate the overlap numerically as

$$O = \int \rho(\mathbf{r}-\mathbf{s}/2)\rho(\mathbf{r}+\mathbf{s}/2) d^3r / \int \rho(\mathbf{r})^2 \, d^3r \quad (5)$$

with the scale factor chosen such that $O=1$ for $\mathbf{s}=0$. [It turned out to be approximately $O \approx 1.66(\mu s)^2 \exp(-2\mu s)$ in the range of interest.] For each of the oxides we consider, with a total of $Z_d$ $d$ electrons per metal ion, there are five majority-spin electrons and $Z_d - 5$ minority-spin electrons. The exchange energy for parallel spin on the two neighboring ions is $5^2 + (Z_d-5)^2$ times $-U_x O$. and for antiparallel ion spins $2\times 5(Z_d - 5)$ times $-U_x O$. The difference gives

$$\Delta E_1 = -(10 - Z_d)^2 U_x O. \quad \text{(overlap exchange)} \tag{6}$$

[$(10-Z_d)^2$ would be replaced by $Z_d^2$ if $Z_d$ were less than five.]

For Mn we find $\mu s=\mu d\sqrt{2}=6.28$ and the overlap $O=0.232\times10^{-3}$, giving an overlap exchange favoring parallel spin for a pair of neighboring Mn ions with $\Delta E_1=-4.5$ meV. It may be surprising that this is nearly as large as the superexchange $\Delta E_1$, but this will not be true for the other oxides. The direct exchange is dominant in any case, and the total $J_1$ becomes 16.4°K.

These values are then in moderate accord on average with the values estimated by Lines and Jones[9], $J_1=10$°K and $J_2 = 11$°K, though our ratio is much smaller. On the other hand, Bloch and Maury[10] found $J_2/J_1= 0.47$, closer to our 0.28. We shall return to the magnetic properties after we make evaluations of $J_1$ and $J_2$ for the other oxides.

### 3. Other Monoxides

We may obtain values for the principal parameters for FeO, CoO, and NiO, exactly as we found them for MnO, and they are listed in Table 1. There are also differences in the analysis because of the additional $Z_d-5$ electrons in minority-spin states. These have an energy $\varepsilon_d^{min}= \varepsilon_d^- +(10-Z_d)U_x$ because the occupation of parallel-spin states is less by $10-Z_d$ than for majority-spin electrons. Similarly, the energy at which minority electron density is introduced from neighbors is higher at $\varepsilon_d^+ = \varepsilon_d^{min}+U_d$ as in MnO. Again in obtaining the total energy we add shifts only for empty states, eliminating those minority states occupied. For this series only $t_g$ minority states are occupied since the three per ion are raised by only $4V_{pd\pi}^2/(\varepsilon_d^+-\varepsilon_p)$ (in second-order perturbation theory) from interaction with the four oxygen $p$ states to which they are coupled, while the two $e_g$ states are raised by $3V_{pd\sigma}^2/(\varepsilon_d^+-\varepsilon_p)$ by the corresponding coupling. There is a difficulty in that if we eliminate a particular $t_g$ state it will affect $\Delta E_1$ differently for different neighbors. We should subtract an average contribution for each $t_g$ dropped, equal to $(V_{pd\sigma}^2 V_{pd\pi}^2+2V_{pd\pi}^4)/3$ in the numerator of Eq. (2) for $\Delta E_1$ and to $2V_{pd\pi}^4/3$ in the numerator of Eq. (3) for $\Delta E_2$.

The situation is a little more complicated for the direct exchange of Eq. (5). A $t_g$ state $xy$ is coupled by $(3V_{dd\sigma}+V_{dd\delta})/4$ to an $xy$ state on an ion in a [110] direction (from the Slater-Koster Tables[11]) which means such a state is $\sqrt{3}/2$ times a σ-oriented state plus 1/2 times a δ-oriented state. That means that the sum of the squared coupling to *all* states on that ion is $(3V_{dd\sigma}^2+V_{dd\delta}^2)/4$ and there are four such neighbors. Similarly there are eight neighbors for which the sum of squared couplings is $(V_{dd\pi}^2+V_{dd\delta}^2)/2$ and the average of the squared coupling for each neighbor is $V_{dd\sigma}^2/4+V_{dd\pi}^2/3+5V_{dd\delta}^2/12 = 0.410V_{dd\sigma}^2$, to be subtracted from the numerator in Eq. (4) once for FeO, twice for CoO, and three times for NiO.

Table 1. Parameters for the monoxides, obtained as for MnO in Section 2. $Z_d$ is the number of $d$ electrons per metal ion. $\varepsilon_d^+=\varepsilon_d+(5-Z_d/2)U_x+U_d$ is the energy at which electrons are added to the $d$ shell from other ions; the oxygen $p$-state energy is $-16.77$ eV in comparison. $J_1$ is the predicted contribution for the nearest-neighbor. $J_2$ is the second-neighbor value.

|  | MnO | FeO | CoO | NiO |
|---|---|---|---|---|
| $Z_d$ | 5 | 6 | 7 | 8 |
| $d(\text{Å})^a$ | 2.22 | 2.16 | 2.12 | 2.08 |
| $V_{pd\sigma}(\text{eV})^a$ | 1.084 | 1.092 | 1.076 | 1.062 |
| $V_{dd\sigma}(d\sqrt{2})(\text{eV})^a$ | 0.202 | 0.189 | 0.173 | 0.159 |
| $U_x(\text{eV})^a$ | 0.78 | 0.76 | 1.02 | 1.60 |
| $U_d(\text{eV})^a$ | 5.6 | 5.9 | 6.3 | 6.5 |
| $\varepsilon_d^+$ (eV) | −7.72 | −9.12 | −9.94 | −10.87 |
| $J_1(°K)_{\text{superexchange}}$ | 2.94° | 5.56° | 8.22° | 14.58° |
| $J_1(°K)_{\text{direct}}$ | 15.52° | 17.80° | 18.56° | 21.62° |
| $J_1(°K)_{\text{overlap}}$ | −2.10° | −1.76° | −1.95° | −2.64° |
| $J_1(°K)_{\text{minority band}}$ | 0° | −4.28° | −5.98° | 0° |
|  |  | Totals |  |  |
| $J_1(°K)_{\text{total}}$ | 16.36° | 17.32° | 18.85° | 33.56° |
| $J_2(°K)_{\text{superexchange}}$ | 4.04° | 9.06° | 16.96° | 43.72° |

a From Ref. 4.

There is an additional effect of these occupied minority states which is related to what is called Stoner band ferromagnetism. The coupling between these minority-spin states will broaden them into bands, of width $W$ which we determine in Appendix A in terms of the second moment $M_2$ of the bands. That $M_2$ is obtained directly from the coupling assuming all ionic moments, and therefore minority spins, are parallel. With partial occupation of the bands only the lower states will be occupied, lowering the energy of the system. If the spins on some fraction of neighbors are not parallel, $W$ is reduced and the energy rises. This can be interpreted as a negative contribution to $J_1$, which we find in Appendix A to be given by

$$\Delta E_1 = -W^2/[54U_d]. \quad \text{(minority band, FeO, CoO)} \tag{7}$$

For MnO the bands are empty and for NiO they are full, and the contribution vanishes.

All of these contributions to $J_1$ and $J_2$ are listed in Table 1. The general rapid growth to the right of the table arises from the factor $1/S^2$ in Eq. (1), with $S$ dropping from 5/2 to 2 to 3/2 to 1 through the series. The $\Delta E_1$ and $\Delta E_2$ are relatively constant, as would be expected from the slow variation of the other parameters which enter, seen in Table 1. The values for overlap exchange use the $O$ from Eq. (5) based upon the $d$-state term value for the metal in question.

## 4. Experimental Consequences

Knowing the origin and value of the various $J$'s in Table 1, we can learn something about the properties which can be described in terms of them. White[12] (also Ref. 1, p. 479) has indicated that all four compounds have Type II ordering of spins in the antiferromagnetic structure. Then (111) planes of the face-centered-cubic metal-ion lattice have parallel spin, but alternate spin from plane to plane. Thus each Mn has six nearest neighbors (in plane) parallel and six (out-of-plane) antiparallel, and all six second neighbors antiparallel. The energy for parallel spin minus that for this antiferromagnetic structure is $6J_1+6J_2$. In the Type I antiferromagnetic structure[12] alternate ions in (100) planes are of opposite spin, giving eight nearest neighbors of each ion of opposite spin, four of the same spin, and all second neighbors of parallel spin for an energy difference of $8J_1$, which would be more stable unless $J_2/J_1>1/3$. Our value for MnO from Table 1, $J_2/J_1= 0.28$ is close to that condition and we shall see that lattice distortions makes up some of the difference of $6J_2-2J_1 = -8.5$ meV per ion. For the other three, we we find Type II much lower in energy, in agreement with experiment.

We look next at the Néel temperature, the mean-field value of which is given for Type II ordering by (Ref. 10, Eq. (6))

$$k_BT_N=4S(S+1)J_2. \tag{8}$$

It depends only upon $J_2$ since half the nearest neighbors are antiparallel in both the ordered Type II and in the random arrangement. Values for $T_N$ were obtained for all four compounds using Eq. (8), and the $J_2$ in Table 1 and are listed in Table 2. These are remarkably close to the experimental values given by Kittel[1], except for NiO, and the trend is correct. They are really closer than we could hope for in view of simplicity of the theory, the 10% corrections to perturbation theory and for self-consistent shifts of the levels, and the uncertainty of the parameters used. Again use of the coupling based upon the Atomic Surface Method of Straub and Harrison[7] would have reduced our estimates to 40% (both the direct and superexchange terms vary as $r_d^6$), far from experiment. It is interesting that reevaluating $T_N$ for Type I ordering (replacing $J_2$ by $2J_1/3-J_2$) predicts 241°K for MnO, 94°K for FeO, and negative values for the other compounds, supporting the assertion[12] that these are all Type II ordering. This value of $T_N$ higher for Type I than Type II MnO reflects our slight overestimate of $J_1/J_2$.

The Néel temperature depended only upon superexchange ($J_2$), proportional to $V_{pdm}^4$, and is expected to vary as $1/d^{16}$. Thus through Eq. (8) we would expect that $(d/T_N)\partial T_N/\partial d = -16$. Bloch and Maury[10] have suggested that experiment is closer to $-10$ but the value is not well established, and they indicate a $(d/J_1)\partial J_1/\partial d = -25$, of much greater magnitude than the $-14$ we shall see is expected for $J_1$.

Table 2. Predicted and experimental Néel temperature $T_N$ and Curie-Weiss temperature $\theta$, based upon mean-field theory and the data in Table 1.

|  | MnO | FeO | CoO | NiO |
|---|---|---|---|---|
| $T_N$ (Eq.(8), °K) | 142 | 218 | 254 | 350 |
| $T_N$ (Exp., °K)[a] | (116) | (198) | (291) | (525) |
| $\theta$ (Eq.(9), °K) | 643 | 525 | 410 | 443 |
| $\theta$ (Exp., °K)[a] | (610) | (570) | (330) | ($\approx$2000) |

a. Ref. 1, p. 481.

We look next at the Curie-Weiss temperature, which determines the susceptibility above the Néel temperature (Ref. 1, p. 479), $\chi = 2C/(T+\theta)$. An expression for the Curie-Weiss temperature for face-centered-cubic lattices was given in Ref. 9, Eq. (2.3),

$$k_B\theta = (12J_1+6J_2)S(S+1)/3. \tag{9}$$

Substituting our values from Table 1 gives $\theta = 643°$ for MnO, larger than the experimental value of 540° which was obtained by Lines and Jones[9] with a fit to the temperature dependence of the magnetic susceptibility at high temperatures; Kittel[1] gives 610°K. The corresponding values for the other oxides are listed in Table 2. Without the minority-band contribution they would show a steady *increase* with increasing $Z_d$, as did $T_N$. Note that this minority-band coupling does not enter the Néel temperature, which depended only upon $J_2$. As with $T_N$, our estimates are surprisingly good, except for NiO and show essentially correct trends.

Since the Curie-Weiss temperature is dominated by direct exchange, and $V_{ddm}$ varies as $1/d^5$ (also $W$ in the minority-spin-band term) we would at first expect $(d/\theta)\partial\theta/\partial d$ to be $-10$. However, our replacement by an exponential would give $(d/\theta)\partial\theta/\partial d = -10\sqrt{2} = -14$ at this second-neighbor spacing. Again, the superexchange contribution is expected to vary as $(d/J)\partial J/\partial d = -16$. We did not find experimental values for the pressure dependence of the Curie-Weiss temperature.

Another property which depends directly on the $J_1$ and $J_2$ is the magnetic susceptibility at the Néel temperature, $\chi(T_N)$, given by Eq. (2.4) in Ref. 9, and proportional to $1/(J_1+J_2)$. With parameters from Table 1 it is 87 emu/g for MnO, again dominated by direct exchange, and very close to the experimental[9] 80 emu/g.

For the Type II antiferromagnetic arrangement in MnO, with parallel (111) planes of alternate spin[12], there is an attraction between adjacent planes, and a repulsion within the planes, relative to the randomized spins. Thus the lattice will contract along this [111] axis, by some factor $1-\varepsilon$, and expand in the lateral directions by a factor $1+\varepsilon/2$. There will be an additional isotropic thermal contraction of the lattice of less interest. The dominant term causing the shear distortion is the direct exchange between nearest neighbors for which the distance to nearest neighbors in adjacent planes changes by $\delta r/r = -\varepsilon/2$ and within the planes by $\delta r/r = \varepsilon/2$. Second neighbor distances do not change to first order in this strain. In the antiferromagnetic state, as opposed to the nonmagnetic state, all six out-of-plane neighbors, rather than three, are antiparallel and no in-plane neighbors, rather than three, are antiparallel, so three nearest neighbors change from parallel to antiparallel in becoming antiferromagnetic. The change in the direct exchange interaction energy for each neighbor pair is $10\sqrt{2}\Delta E_1 \delta r/r$, half associated with each ion, if it is antiparallel rather than parallel, so the energy per ion changes by $-10\sqrt{2}\Delta E_1(3\varepsilon/4)$ per ion if the crystal becomes antiferromagneticly ordered. There is also elastic energy for the three shears, $e_4=e_5=e_6=\varepsilon$, of $(3/2)c_{44}\varepsilon^2 2d^3$ per ion with[10] $c_{44} = 0.57$ megabars = $0.36$ eV/Å$^3$. We may minimize the total to obtain the fractional reduction in length along the [111] direction of

$$\varepsilon = 10\sqrt{2}\Delta E_1/8c_{44}d^3 = 0.015 \qquad (10)$$

due to direct exchange. The corresponding contribution from superexchange and overlap exchange increase this to a total fractional length change along [111] of $\Delta L/L = -\varepsilon = -0.0167$ and expansion in the lateral directions of 0.0083. These are in rough accord with the –0.013 and 0.004 obtained from experiment by Bloch and Maury (Ref. 10, Fig. 5).

Our estimate corresponds to a change in energy for the crystal of –3.3 meV per ion, favoring this Type II planar spin arrangement over the Type I cubic alternative which has no corresponding shear relaxation. This makes up only part of the difference of $6J_2-2J_1 = -8.5$ meV which we found at the beginning of this section. These distortions are additionally important since, as pointed out by Bloch and Maury[10], they make the antiferromagnetic transition first order. We would seem to have all the parameters needed for a mean-field theory of the $\Delta L/L$ as a function of temperature, and its discontinuity at $T_N$, but we have not carried it out.

CuO does not form in the rock-salt structure, but it is straightforward to evaluate the nearest-neighbor $J$ for $CuO_2$ planes in the cuprates using parameters obtained exactly as for the other compounds. The Cu-O-Cu lie in a straight line as for $J_2$ in the rock-salt structure. A single $x^2-y^2$ is empty on each Cu ion, coupled to O neighbors by[11] $\sqrt{3}V_{pd\sigma}/2$ ($V_{pd\sigma}=-1.35$ eV, slightly larger than given in Ref. 13 because of the use of MTO $r_d$ values) giving a coupling to a neighboring Cu as

$$\Delta E_1 = 4\frac{9V_{pd\sigma}^4}{16(\varepsilon_d^+ - \varepsilon_p)^2(\varepsilon_d^+ - \varepsilon_d^-)} \quad \text{(cuprate)} \qquad (11)$$

This corresponds to a $J = 1220°$ to be compared with the experimental[14] 1500°K.

Similarly, CrO seems not to form in the rock-salt structure, but we consider the compound $LiCrO_2$ discussed by Mazin[15]. The structure is similar to that for MnO but with alternate (111) metal-ion planes replaced by a sprinkling of Li ions. With three electrons in $d$ states it has a filled majority $t_g$ shell, and other states empty. In the distorted structure, with a distance between Cr ions of $a = 2.90$ Å, $V_{dd\sigma} = 0.472$ eV. Direct exchange gives a $\Delta E_1=0.272$ eV from Eq. (4), corresponding to a $J_1$ of 30 meV, and superexchange ($d=1.97$ Å), minus a small overlap exchange, would raise it 9meV. The total is somewhat above the experimental value listed by Mazin[15] of 20 meV. The $J_2$ of MnO does not arise in this structure.

The predictions have generally been close enough to experiment to suggest that the theory is essentially correct and the discrepancies, where they exist, could come from using the simplest theory of the properties. The analysis has given a clear picture of the various contributions, and their relative magnitudes, which seems not to have previously available. There is uncertainty in the choice of parameters, particularly for $r_d$ and this study strongly supports use of the MTO values. This had not been so clear in earlier studies. We recalculated the dielectric susceptibilities and transverse charges from Ref. 8 using these MTO values. These properties are essentially proportional to $V_{pdm}^2$, rather than $V_{pdm}^4$, so the changes are much smaller, and the differences were half as big for $4d$ and $5d$ rows as for $3d$ systems. For the alkali halides, where we added contributions from alkali-metal $d$ states with parameters extrapolated from the transition metals, the scaling up of the $r_d$ values improved agreement for the susceptibilities but changes were mixed for the effective transverse charges. Corrections for the alkaline earths were also mixed. For the one transition-metal system treated, $SrTiO_3$, use of the MTO values improved agreement in more cases than not, but not enough to make a convincing case for MTO values. There is also limited meaning to use of a single value for $U_d$, as seen in Appendix B.

## 5. Acknowledgement

The author is indebted to Professor Robert M. White for raising the questions addressed here, and to him and to G. Deutscher for helpful discussions.

## Appendix A  Correlated Electrons

Our starting states for the analysis in this paper, electrons localized on individual ions, depends directly on electron correlations. These can be considered more closely by examining a single pair of electrons in single coupled orbitals, treated, for example, in Ref. 4, 593ff. Two orbitals of energy $\varepsilon_s$ are coupled by a $V$, which could be a direct coupling like our $V_{ddm}$ or a second-order coupling such as our $V = V_{pdm}^2/(\varepsilon_d^+ - \varepsilon_p)$ for superexchange. There is an extra energy $U$ if both electrons are on the same atom, as in Appendix B. Of the six two-electron states (e. g., $c_{1+}^\dagger c_{2-}^\dagger |0>$ with the numbers labeling atoms and the ± indicating spin ) only four have antiparallel spin and, with orbitals of the same energy, two symmetric combinations, giving a quadratic equation for the ground-state energy. It yields an exact energy for the two-electron state,

$$E_{TOT} = 2\varepsilon_s + U/2 - ((U/2)^2 + (2V)^2)^{1/2}. \tag{A1}$$

For small $U$ it gives two electrons in bonds at $\varepsilon_s - |V|$ plus $U/2$ for the 50% chance the two electrons are on the same site. For large $U$ the energy is $2\varepsilon_s - 4V^2/U$, while in a one-electron theory the shift would have been only $-2V^2/U$ for the two electrons. This feature results from each of the two terms in the symmetrized combinations of two-electron states being coupled to both of the terms in the second symmetrized combination. The fully symmetrized ground state has lower energy than the sum of the two one-electron states obtained in perturbation theory. This smaller value was also found (Ref. 4, p. 495) by adding a $U$ to a one-electron calculation, allowing the two states to disproportionate, a kind of LDA plus $U$. It seemed clear that we should use the larger energy shift in the analysis here.

We may also reexamine the extension we made (Ref. 4, 598ff) of Eq. (A1) to $d$ and $f$-shell metals. There we were interested in other properties, but here we use it to calculate the Stoner-like effects of the minority spin-bands in FeO and CoO. In Ref. 4 we replaced the band density of states by a constant ($10/W$ per atom for $d$ bands) over a width $W$. One readily finds that such a density of states has a second moment of $M_2 = W^2/12$, allowing us to determine $W$ directly from the Hamiltonian matrix from which the bands would be calculated. We also readily find that occupying such a density of states with $Z_d$ electrons per atom gains an energy $-Z_d(1-Z_d/10)W/2$ per atom. In Ref. 4 we generalized Eq. (A1) to the gain in energy per atom due to coupling as

$$E_{TOT} = -\frac{Z_d(1-Z_d/10)}{2}\left[\sqrt{W^2 + \zeta^2 U_d^2} - \zeta U_d\right], \tag{A2}$$

with $\zeta=1$. This gives the appropriate result for $U_d=0$, for all $\zeta$, and we may now choose $\zeta$ such that it gives the correct result for large $U_d$. That result for $Z_d$ occupied states per atom coupled to $10-Z_d$ empty states on each nearest neighbor is $-2Z_d(1-Z_d/10)M_2/U_d$ since $M_2$ is the sum of squared couplings of all states on an atom to all of the states to which they are coupled, divided by the number of states per atom, here reduced by the fraction of neighboring states empty. The leading factor of two corresponds to the use of $-4V^2/U$ rather than $-2V^2/U$ as just discussed. Expanding the square root in Eq. (A2) and equating it to this perturbation-theory gain, using $M_2 = W^2/12$, we see that $\zeta = 3/2$. In Ref. 4 we assumed $\zeta = 1$.

For our treatment of the minority $t_g$ band with $Z$ electrons in three, rather than ten, states we have the energy per metal ion of

$$E_{TOT} = -\frac{Z(1-Z/3)}{2}\left[\sqrt{W^2 + \frac{9U_d^2}{4}} - \frac{3U_d}{2}\right] \quad (A3)$$

The leading factor is -1/3 for FeO and CoO, and zero for MnO and NiO. To second order in $W$ this becomes $-W^2/9U_d$ for FeO and CoO.

The second moment of the band, $M_2$, may be obtained from the $3N$-by-$3N$ Hamiltonian matrix for $N$ metallic atoms (e. g., Ref. 4, p. 560), or the corresponding three-by-three matrix. The squared couplings for $t_g$ orbitals are closely related to the squared couplings we obtained for direct exchange in Section 3, where we noted that a $t_g$ state $xy$ is coupled by $(3V_{dd\sigma}+V_{dd\delta})/4$ to an $xy$ state on a neighbor in a [110] direction. However, it is not coupled to the other $t_g$ states on that atom so the squared coupling is $(3V_{dd\sigma}+V_{dd\delta})^2/16$, to be added for four atoms, quite a bit smaller than the sum of squared coupling to *all* orbitals on the second atom. On the other hand, for the other eight neighbors, all states coupled to the $xy$ state are $t_g$ states so the sum of the squared couplings is the same. The final sum of squared $t_g$ couplings per neighbor is $0.366V_{dd\sigma}^2$, rather than the $0.410V_{dd\sigma}^2$ we had before. For twelve neighbors we would obtain $W^2 = 52.70V_{dd\sigma}^2$ if all of the neighboring $t_g$ orbitals were of the same spin as the one under consideration (and therefore coupled), half of that if half were of the same spin.

We see that the lowering of energy in Eq. (A3) is favoring parallel spins, a negative contribution to $J_1$. For a fraction $f$ of the neighboring spins parallel we have an energy per ion to second order of $-W^2 f/9U_d$, which we may equate to an energy $4S^2 J_1 12f/2$, from Eq. (1), for 12 neighbors, counting half for each shared ion.

$$J_1 = -W^2/[216S^2 U_d]. \quad (A4)$$

We may evaluate this immediately from Table 1 and obtain a $J_1$(FeO) = $-4.28°$K and $J_1$(CoO) = $-5.98°$K, which are listed also in Table 1. These give corrections to the θ of Table 2 of $-103°$ for FeO and $-90°$ for CoO, both of the right general magnitude and included in Table 2.

The actinides which were treated this way in Ref.16 were electrically conducting, but they would have been conducting due to the *s* and *d* bands even without the highly correlated *f* levels. In Ref. 16 we assumed a density of electronic states at the Fermi energy based upon an *f*-band width $W^2/(W^2+\zeta^2 U_d^2)^{1/2}$ with $\zeta=1$ to estimate the electronic specific heat and Pauli susceptibility, and in Ref. 17 did the same for transition metals. It would seem that the $\zeta=3/2$ we used here for the total energy should apply also to the density of states [using the form with the square root for the band width since in these systems *W* was not small compared to $U_d$ and predicted densities of states were increased by around 30%]. Using it for the transition metals generally improved the estimates, but the large fluctuations of the density of states, not present in the assumed constant density of states, made it difficult to compare. Using $\zeta=3/2$ for the actinides would have improved predictions for the susceptibility but not for the specific heat. Couplings for these analyses were based upon $r_f$ values[4] from MTO theory.

For FeO and CoO it would not be surprising if the minority-spin electrons had been conducting, but neither is it surprising that they were not. The coupling between neighboring levels, if the spins are aligned, gives shifts of the order of the coupling and electrons in the lower levels may or may not find empty levels close enough in energy and position to allow them to flow. Our summing of occupied levels over a distribution of width *W* of about one eV (for half alignment) does not require metallic conductivity, though one usually associates Stoner corrections with metals.

## Appendix B  The Coulomb $U_d$

There is uncertainty in just what values are appropriate for $U_d$. The values we use were obtained by Straub (Ref. 4, p. 561) from Hartree-Fock calculations on the free atom to obtain the shift in a *d*-state energy when a second electron was transferred from an *s* state to a *d* state. This could alternatively be obtained from Moore's spectroscopic tables[18]. We can for example compare the energy in going from the doubly-ionized to triply-ionized Mn as $d^5$ to $d^4$ and as $d^4s$ to $d^3s$, the former taking less energy by $U_d$. From the tables we learn, by subtracting the term value for $d^5$ from that for $d^4s$ in the doubly-ionized atom, that the second starting state is higher by 7.81 eV. [There was some spread of values which could be chosen depending upon the total angular quantum number *J* of the states in question but not on the scale of eV as long as the total spin *S* was the same.] We similarly learn that the second final state is higher than the first by 13.88 eV, which gives us $U_d$= 6.07 eV, the difference. We could instead go from triply to quadruply ionized states, comparing $d^4$ to $d^3$ with $d^3s$ to $d^2s$ yielding $U_d$= 22.06−13.88 eV = 8.18 eV.  We could not go from singly ionized to doubly ionized states since no term value was available for a $d^3s^2$ state.  The difference between 6.07 and 8.18 eV indicates that the spectra are not describable by a single $U_d$ and our Hartree-Fock value of 5.6 eV in Table 1 is not unreasonable

    Using this $U_d$ in this paper as the energy shift when an electron is transferred from a *neighboring* ion assumes that the average distance of an atomic *s* electron from the nucleus is equal to the distance to the neighbors, an approximation which has often worked well[4]. As a check on the use of this $U_d$ for our calculations we again use Moore's tables[18] to obtain the energy required to transfer a *d* electron from a *d* state on one Mn ion to a *d* state on a neighboring Mn, proceeding step by step. We first obtain the value of *U* for widely separated ions , using two Mn ions (singly ionized) in the starting configuration $d^6$, one going to $d^5$ and the other to $d^7$. Indeed this energy is higher by the Coulomb repulsion *U* between two *d* electrons, but by putting the two minority electrons on the same ion  we also gain an exchange energy $U_x$ = 0.78 eV, which we had earlier[4] obtained from the atomic spectra, so the energy change in going $d^6$, $d^6$ to $d^5$,$d^7$ is $U_d$−$U_x$.

    In the first step both ions go from $d^6$ to the ground state of the singly ionized Mn, $d^5s$, the energy dropping by −1.78 eV for each ion. Removing an *s* electron from one Mn  then takes the ionization potential of the $Mn^+$ equal to  15.64 eV (leaving it as $d^5$).  Adding it to the other gains back the ionization energy of  $Mn^0$, 7.43 eV (leaving it as $d^5s^2$). Another 6.40 eV is required, obtained by subtracting term values for the $Mn^0$, to take the $d^5s^2$ ion to $d^7$, for a total cost of  $U_d$−$U_x$= 11.05 eV for the widely separated pair, two $d^6$ ions to a $d^5$ and a $d^7$.

    If these ions were a distance  *r*  apart, this energy would be reduced by −$e^2$/*r*. On the other hand, in the solid it would seem appropriate to use the Madelung potential in a crystal with electrons shifted between all (111) planes as in the antiferromagnetic arrangement of MnO.  That Madelung constant seemed not to be known, but we calculated it using the simple procedure described in Ref. 19. [Note in the case of (111) planes, the sum over one eighth of space for *x*, *y*, and *z* all greater than zero or for all less than zero (each giving $0.351e^2/d$) was much greater than for each of  the other six eighths where only one coordinate, or two coordinates,  was less than zero (each giving $0.0282e^2/d$).] The net result was $0.871e^2/d$, with  *d* the Mn-O nearest-neighbor distance.  Note that adding the positive charge of +2 on each Mn and −2 on each O does not affect this value. With *d*=2.22 Å, this reduces the $U_d$−$U_x$ by 5.65 eV to 5.40 eV, giving $U_d$=6.18 eV, between

the two intra-atomic $U_d$ values we obtained from the spectra, and close enough to the 5.6 eV we used as not to significantly affect the comparison with experiment.

Note that the 5.6 eV came from a Hartree-Fock calculation and the 6.18 eV from experimental spectra. There are similar differences between the magnitude of the Hartree-Fock term values ($\varepsilon_s = -6.84$ eV for Mn) and experimental ionization energies (7.43 eV for Mn). There would be similar differences with values obtained with local-density theory, or LDA plus $U$. The atoms in the bulk antiferromagnetic crystal do not of course have these shifted charges, just majority spins in opposite directions. The Madelung constant enters the calculation because the excited state which enters the perturbation theory is an added minority-electron state.